\documentclass [12pt,a4paper]{article}
\usepackage{bbm}
\usepackage{amsfonts}
\usepackage{mathrsfs}
\usepackage {amssymb}
\usepackage {amsmath}
\usepackage{amsthm}
\usepackage{latexsym}
\usepackage {amssymb}
\usepackage {amsmath}
\usepackage{color}

\makeatletter
 \@addtoreset{equation}{section}
 
\makeatother

\usepackage{lastpage}
\usepackage{epsfig}

\begin{document}
\begin{center}
\textbf{\Large{On the 2-adic complexity of a class of binary sequences of period $4p$ with optimal autocorrelation magnitude} } {\footnote { Minghui Yang was supported by NSFC under Grant 11701553. Lulu Zhang was supported by the National Natural Science Foundation of China (NSFC) under Grant 11601350.  Keqin Feng was supported by NSFC under Grant 11571007 and 11471178.

Minghui Yang is with the State Key Laboratory of Information Security, Institute of Information Engineering, Chinese Academy of Sciences, Beijing 100093, China, email: (yangminghui6688@163.com).

Lulu Zhang is with Capital Normal University, Beijing 100048, China, (e-mail: 840375411@qq.com).

Keqin Feng is with the department of Mathematical Sciences, Tsinghua University, Beijing, 100084, China, email: (fengkq@tsinghua.edu.cn). }
}

\end{center}

\begin{center}
\small Minghui Yang,  Lulu Zhang, Keqin Feng
\end{center}

\begin{center}
\small
\end{center}

%-------------------------------------------------------------------------

\noindent\textbf{Abstract}-Via interleaving Ding-Helleseth-Lam sequences, a class of binary sequences of period $4p$ with optimal autocorrelation magnitude was constructed in \cite{W. Su}. Later, Fan showed that the linear complexity of this class of sequences is quite good \cite{C. Fan}. Recently, Sun et al. determined the upper and lower bounds of the 2-adic complexity of such sequences \cite{Y. Sun3}. We determine the exact value of the 2-adic complexity of this class of sequences. The results show that the 2-adic complexity of this class of binary sequences is close to the maximum.

\noindent\textbf{keywords}-2-adic complexity, optimal autocorrelation magnitude, binary sequences.

\section{Introduction}
\label{}
Sequences with good randomness such as long period, low autocorrelation and large linear complexity are widely used in cryptography, communication, etc. Feedback with carry shift registers (FCSRs) are a class of nonlinear pseudo random sequence generators. Due to the rational approximation algorithm \cite{A. Klapper}, 2-adic complexity has become an important security criteria. Hence, it is interesting to investigate the 2-adic complexity of some well-known sequences with optimal autocorrelation and large linear complexity.

The autocorrelation function of binary sequence $s=(s_0, s_1, \ldots, s_{N-1})$ with period $N$ is defined by
$$C_s(\tau)=\sum_{i=0}^{N-1}(-1)^{s_i+s_{i+\tau}}, \ \ \ \tau\in\mathbb{Z}/N\mathbb{Z}.$$
A sequence $s$ with period $N$ is called an optimal autocorrelation sequence \cite{K. Arasu} if for any $\tau\neq 0$,

(1) $C_s(\tau)=-1$ for $N \equiv 3\pmod 4$; or

(2) $C_s(\tau)\in\{1,-3\}$ for $N \equiv 1\pmod 4$; or

(3) $C_s(\tau)\in\{2,-2\}$ for $N \equiv 2\pmod 4$; or

(4) $C_s(\tau)=0$ for $N \equiv 0\pmod 4$.

Up to equivalence, the only known binary sequence in Type (4) is (0, 0, 0, 1). Hence, for a sequence with period $N\equiv0\pmod 4$, it is natural to consider the case $C_s(\tau)\in\{0,\pm4\}$. When $\tau$ ranges from 1 to $N-1$, $s$ is referred to as a sequence with optimal autocorrelation value if $C_s(\tau)\in\{0, -4\}$ or $\{0, 4\}$ \cite{X. H. Tang}, and $s$ is referred to as a sequence with optimal autocorrelation magnitude if $C_s(\tau)\in\{0,\pm4\}$ \cite{N. Y. Yu}.

Interleaved operator that was originally presented by Gong \cite{G. Gong} is a powerful tool to construct sequences with optimal autocorrelation and large period.

Let $s^t=(s_0^t, s_1^t, \ldots, s_{N-1}^t)$ be a binary sequence of period $N$, where $0 \leq t\leq M-1$. An $N \times M$ matrix is obtained from these $M$ binary sequences and given by
\begin{equation*}U=
\left(
\begin{array}{cccc}
s_0^0 & s_0^1 & \cdots & s_0^{M-1}\\
s_1^0 & s_1^1 & \cdots & s_1^{M-1}\\
\vdots & \vdots &\ddots &  \vdots\\
s_{N-1}^0 & s_{N-1}^1& \cdots & s_{N-1}^{M-1}
\end{array}
\right).
\end{equation*}
An interleaved sequence $u=(u_h)$ of period $MN$ is obtained by concatenating the successive rows and defined by
$$u_{i M+j}=U_{i, j}, 0 \leq i <N, 0 \leq j <M.$$
The sequence $u$ is denoted by
$$u=I(s^0, s^1, \ldots, s^{M-1})$$
for simplicity.

Recently, using Ding-Helleseth-Lam sequences defined in \cite{C. Ding} and a binary sequence $\mathbf{b}=(b(0), b(1), b(2), b(3))$ with $b(0)=b(2)$, $b(1)=$
$b(3)$, Su et al. \cite{W. Su} constructed  a new class of binary sequences of period $4p$ with optimal autocorrelation magnitude by interleaving operator.  Later, Fan \cite{C. Fan} proved that the linear complexity of these sequences is close to the maximum.

The 2-adic complexity of binary sequences with good autocorrelation has not been studied so fully as the linear complexity.  The 2-adic complexity of sequences in Type (1) was studied in \cite{H. G. Hu, T. Tian, H. Xiong 1}. Very recently, the 2-adic complexity of Ding-Helleseth-Martinsen sequence with period $2p$ in Type (3) was determined in \cite{L. Zhang} by using ``Gauss periods" and ``Gauss sum" on finite field $\mathbb{F}_q$ valued in the ring $\mathbb{Z}_{2^{2p}-1}$. The 2-adic complexity of some other sequences with good autocorrelation was studied in \cite{R. Hofer, Y. Sun1, Y. Sun2, Y. Sun3, Z. B. Xiao, H. Xiong 2}. Specially, Sun et al. \cite{Y. Sun3} presented the 2-adic complexity of the upper and lower bounds of interleaved sequence $u$ constructed from \cite{W. Su} when $\mathbf{b}=(b(0), b(1), b(2), b(3))=(0, 1, 0, 1)$ by using Hu's method \cite{H. G. Hu} that associates with the autocorrelation function. In the conclusion of their paper, they guessed the upper bound can be arrived which means $\gcd(u(2), 2^{2p}+1)=5$  where $u(x)=u_0+u_1x+\cdots +u_{4p-1}x^{4p-1}$.

In this paper, we prove the guess in \cite{Y. Sun3} is right inspired by \cite{L. Zhang}. Furthermore, we determine the exact value of the 2-adic complexity of other interleaved sequences constructed in \cite{W. Su} with binary sequence $\mathbf{b}=(b(0), b(1), b(2), b(3))$ satisfying $b(0)=b(2)$, $b(1)=b(3)$.

\section{Preliminaries}

\ \ \  \ In this section, we will introduce some notations and well-known results.

From now on, we adopt the following notation without special explanation.

$\bullet$ Let $u=(u_0, u_1, \ldots, u_{N-1})$ be a binary sequence of period $N$. The set
$$B_u=\{t\in\mathbb{Z}_N: u_t=1\}$$ is called the support of $u$.

$\bullet$  $U(x)=\sum_{i=0}^{N-1}u_ix^i\in\mathbb{Z}[x]$, $T(x)=\sum_{i=0}^{N-1}(-1)^{u_i}x^i$.

$\bullet$ $u+1$ is defined by $u+1=(u_0+1, u_1+1, \ldots, u_{N-1}+1)$.

$\bullet$ The cyclic left shift operator of $u$ is defined by $$L^e(u)=(u_e, u_{e+1}, \ldots, u_{N-1}, u_0, \ldots, u_{e-1}),$$ where $0 \leq e\leq N-1.$

$\bullet$ $d$ is a positive integer satisfying $4d \equiv 1\pmod p$.

$\bullet$ Let $g$ be a primitive root of $p$. Define $D_j=\{g^{j+4i}: 0 \leq i \leq \frac{p-1}{4}-1\}$ for $0 \leq j \leq 3$.

$\bullet$ Let $s^1, s^2, s^3$ be the Ding-Helleseth-Lam sequences of period $p$ with supports $D_0\cup D_1$, $D_0\cup D_3$, $D_1\cup D_2$, respectively, where $p=4f+1=x^2+4y^2$ is a prime number, $f$ is odd and $y=\pm 1$.

$\bullet$ $``\gcd"$ denotes the greatest common divisor.

By using the interleaved operator, Su, Yang and Fan \cite{W. Su} designed binary sequence of period $4p$ with autocorrelation magnitude. The following result was given by them.

\textbf{Lemma 2.1}(\cite{W. Su}) Let $\mathbf{b}=(b(0), b(1), b(2), b(3))$ be a binary sequence with $b(0)=b(2), b(1)=b(3).$ Then the binary sequence of period $4p$ constructed by
\begin{equation*}
u=I(s^3+b(0), L^d(s^2)+b(1), L^{2d}(s^1)+b(2), L^{3d}(s^1)+b(3))
\end{equation*}
is optimal with respect to the autocorrelation magnitude, i.e., $C_u(\tau)\in\{0, \pm4\}$ for all $0<\tau<4p$.

 Assume that
$$\frac{U(2)}{2^N-1}=\frac{\sum_{i=0}^{N-1}u_i2^i}{2^N-1}=\frac{a}{e}, 0 \leq a \leq e, \gcd(a, e)=1.$$
Then the 2-adic complexity $\Phi_2(u)$ \cite{A. Klapper} is defined by $\log_2\frac{2^N-1}{\gcd(2^N-1, U(2))}.$ Therefore, determining $\Phi_2(u)$ is equivalent to determining $\gcd(2^N-1, U(2))$.

\section{Main result}

In this section, we study the 2-adic complexity of the binary sequence $u$ with optimal autocorrelation magnitude in Lemma 2.1. Firstly, for a sequence $u$ constructed with  $\mathbf{b}=(b(0), b(1), b(2), b(3))=(0, 1, 0, 1)$, we prove that the guess $\gcd(U(2), 2^{2p}+1)=5$ proposed by Sun et al. in \cite{Y. Sun3} is right. Then we determine the exact value of the 2-adic complexity of the sequence $u$ defined in Lemma 2.1.

The following lemma is useful in our paper.

\textbf{Lemma 3.1}
$(\sum_{i\in\mathbb{F}_p^{\ast}}(\frac{i}{p})2^{4i})^2 \equiv p\pmod {\frac{2^{2p}+1}{5}}$, where $(\frac{i}{p})$ is the Legendre symbol  defined by
\begin{align*}
(\frac{i}{p})
 = \left\{ \begin{array}{ll}
0, & \textrm{if $i\equiv 0 \pmod p$},\\
1, & \textrm{if $i \not \equiv 0 \pmod p$ and $i$ is the square of an element of $\mathbb{F}_p^{\ast}$},\\
-1, & \textrm{otherwise}.
\end{array} \right.
\end{align*}

\begin{proof}
Since $(\frac{i}{p})$  is a multiplicative character, we have
\begin{align}
(\sum_{i\in \mathbb{F}_p^{\ast}}(\frac{i}{p})2^{4i})^2& = \sum_{a, b=1}^{p-1}(\frac{ab}{p})2^{4(a+b)}\notag\\
                                  &\equiv\sum_{a, c=1}^{p-1}(\frac{a^2c}{p})2^{4a(1+c)} \ \textrm{(let $b=ac$)}\notag\\
                                  &\equiv\sum_{a, c=1}^{p-1}(\frac{c}{p})2^{4a(1+c)}\notag\\
                                  &\equiv\sum_{c=1}^{p-1}(\frac{c}{p})\sum_{a=1}^{p-1}2^{4a(1+c)}\pmod{2^{4p}-1}
\end{align}
Since $p\equiv 1\pmod 4$, we have $(\frac{-1}{p})=1$ and then the contribution of $c=p-1$ to the right hand side of (3.1) is
$$\sum_{a=1}^{p-1}2^{4ap} \equiv p-1 \bmod (2^{4p}-1).$$
From $(\frac{-1}{p})=1$ we know $\sum_{c=1}^{p-2}(\frac{c}{p})=-1$ and then
\begin{align*}
(\sum_{i\in \mathbb{F}_p^{\ast}}(\frac{i}{p})2^{4i})^2  & \equiv p-1+\sum_{c=1}^{p-2}(\frac{c}{p})(-1+\sum_{a=0}^{p-1}2^{4a(1+c)}) \pmod{2^{4p}-1}\\
                                  & \equiv p-1-\sum_{c=1}^{p-2}(\frac{c}{p})+\sum_{c=1}^{p-2}(\frac{c}{p})\sum_{a=0}^{p-1}2^{4a(1+c)} \pmod{2^{4p}-1}\\
                                  &\equiv\ p-\sum_{a=0}^{p-1}2^{4a} \pmod{2^{4p}-1} \\
                                  &\equiv p\pmod{\frac{2^{2p}+1}{5}}.
\end{align*}

\end{proof}

$\mathbf{Remark:}$ The proof of Lemma 3.1 is similar to Lemma 2.4(1) in \cite{L. Zhang}. For the completeness of the paper, we give a proof here.

Let $\mathbf{\overline{b}}=(\overline{b(0)}, \overline{b(1)}, \overline{b(2)}, \overline{b(3)})$ be the complement of $\mathbf{b}=(b(0), b(1), b(2), b(3)).$ Let $\overline{u}$ and $u$ be constructed with $\mathbf{\overline{\mathbf{b}}}$ and $\mathbf{\mathbf{b}}$ respectively in Lemma 2.1. Then $\overline{u}$ is the complement of $u$, i.e., $\overline{u}=u+1$. Therefore we have
\begin{align*}
\overline{U}(2)=&u_0+1+(u_1+1)\cdot 2+\cdots+(u_{N-1}+1)2^{N-1}\\
                                   =& U(2)+2^N-1\equiv U(2)\pmod {2^N-1}.
\end{align*}
Thus $\gcd(\overline{U}(2), 2^N-1)=\gcd(U(2), 2^N-1)$ and then $\Phi_2(\overline{U})=\Phi_2(U)$.

There are four cases for $\mathbf{b}$ satisfying $b(0)=b(2), b(1)=b(3)$, i.e., $\mathbf{b}=(b(0), b(1), b(2), \\ b(3))=(1, 0, 1, 0), (0, 1, 0, 1),(0, 0, 0, 0), (1, 1, 1, 1).$ In order to determine the 2-adic complexity of the sequence with optimal autocorrelation magnitude in Lemma 2.1, we only need to consider the 2-adic complexity of $u'$ and $u''$ constructed with $\mathbf{b}=(b(0), b(1), b(2), b(3))=(0, 1, 0, 1)$ and (0, 0, 0, 0), respectively.

In the following, we will denote by $u'$ and $u''$ the sequence constructed with $\mathbf{b}=(b(0), b(1), b(2), b(3))=(0, 1, 0, 1)$, and (0, 0, 0, 0) in Lemma 2.1, respectively. Denote $U(x), T(x)$ by $U'(x), T'(x)$ and $U''(x), T''(x)$ for $u'$ and $u''$, respectively.

 We determine the 2-adic complexity of $u'$. The following two lemmas have been proved by Sun et al. in \cite{Y. Sun3}.

\textbf{Lemma 3.2} (\cite{Y. Sun3}) Let the symbols be the same as before. Then

 \begin{align*}
& U'(2)T'{(2^{-1})}\\
& \equiv2\left[ \frac{2^{4p}-1}{2^{4}-1}+(2^{2p}+1)(2^p-1)-2^p(2^{2p}-1)y\sum_{i\in\mathbb{F}_p^{\ast}}(\frac{i}{p})2^{4i}-p \right]\pmod {2^{4p}-1}.
\end{align*}

\textbf{Lemma 3.3}(\cite{Y. Sun3}) $\gcd(U'(2), 2^{2p}-1)=1$ and $5|\gcd(U'(2), 2^{2p}+1)$.

The following theorem shows that the guess of Sun et al. in \cite{Y. Sun3} is right.

\textbf{Theorem 3.4}
For the sequence $u'$, we have $\gcd(U'(2), 2^{2p}+1)=5$.
\begin{proof}
(i) Assume that $p \neq 5.$

From Lemma 3.2 we get
$$U'(2)T'{(2^{-1})}\equiv2  [ -2^p(2^{2p}-1)y\sum_{i\in\mathbb{F}_p^{\ast}}(\frac{i}{p})2^{4i}-p]\pmod{\frac{2^{2p}+1}{5}}.$$
Suppose that $U'(2)$ and $\frac{2^{2p}+1}{5}$ have a common prime factor $l$. Then
\begin{align*}
0\equiv U'(2)T'(2^{-1})& \equiv  2[-2^p(2^{2p}-1)y\sum_{i\in\mathbb{F}_p^{\ast}}(\frac{i}{p})2^{4i}-p]\pmod l\\
                                  & \equiv 2[-2^p(-2)y\sum_{i\in\mathbb{F}_p^{\ast}}(\frac{i}{p})2^{4i}-p]\pmod l.
\end{align*}
Therefore $2^{p+1}y\sum_{i\in\mathbb{F}_p^{\ast}}(\frac{i}{p})2^{4i}-p\equiv 0\pmod l$. From $y=\pm 1$ we get $$2^{2p+2}(\sum_{i\in\mathbb{F}_p^{\ast}}(\frac{i}{p})2^{4i})^2-p^2\equiv 0\pmod l.$$
From Lemma 3.1 we get $0\equiv 2^{2p+2}p-p^2\equiv -4p-p^2\pmod l$ which implies that $l=p$ or $l|p+4$. If $l=p$, by Fermat's Little Theorem,  we get $0\equiv 2^{2p}+1\equiv 5\pmod p$ which contradicts to the assumption $p\neq 5$. If $l|p+4$, from $2^{2p}\equiv -1\pmod l$ we know that $l\neq 3$ and the order $D$ of $2\pmod l$ is 4 or $4p$. From $D|l-1$ and $l|p+4$ we know that $D\neq 4p.$ From $p\neq 5$ and $$\frac{2^{2p}+1}{5}=[1+(-2^2)+\cdots+(-2^2)^{p-2}+(-2^2)^{p-1}]\equiv p\pmod 5,$$
 we have $\gcd(\frac{2^{2p}+1}{5}, 5)=1$  which implies that $l\neq5.$ If $D=4$, then $0\equiv 2^4-1\equiv 15\pmod l$ which contradicts to $l\neq 3, 5$. Therefore $\gcd(U'(2), \frac{2^{2p}+1}{5})=1.$ From Lemma 3.3, we get
$$\gcd(U'(2), 2^{2p}+1)=\gcd(U'(2), \frac{2^{2p}+1}{5})\gcd(U'(2),5)=5.$$
(ii) Assume that $p=5$.

From $\mathbb{F}_5^{\ast}=\langle2\rangle$, we know the cyclotomic classes of order 4 in $\mathbb{F}_5$ are
$D_0=\langle1\rangle$, $D_1=\langle2\rangle$, $D_2=\langle4\rangle$, $D_3=\langle3\rangle$. Since $s^2$ is a binary sequence with support $B_{s^2}=D_0\cup D_3$, we have $B_{L^d(s^2)}=(D_0\cup D_3)-d$ and $B_{L^d(s^2)+1}=(D_1\cup D_2\cup \{0\})-d$. From $4d\equiv 1\pmod p$, we have $-d\equiv \frac{p-1}{4}\pmod p.$ Then by the definition of $u'$, we get

\begin{align*}
 U'(2)& = \sum_{i\in D_1\cup D_2}2^{4i}+\sum_{i\in\{\frac{p-1}{4}\}\cup(D_1\cup D_2)+\frac{p-1}{4}}2^{4i+1}\\& \ \ \
   +\sum_{i\in (D_0\cup D_1)+\frac{p-1}{2}}2^{4i+2}+\sum_{i\in\{\frac{3(p-1)}{4}\}\cup ((D_2\cup D_3)+\frac{3(p-1)}{4})}2^{4i+3}\\
                                  & = \sum_{i\in\{2,4\}}2^{4i}+\sum_{i\in\{1, 3, 5\}}2^{4i+1}+\sum_{i\in\{3, 4\}}2^{4i+2}+\sum_{i\in\{3, 2, 1\}}2^{4i+3}\\
                                  & = 2484640\\
                                  & \equiv\left\{ \begin{array}{ll}
15, \pmod {25},\\
40, \pmod {41}.
\end{array} \right.
\end{align*}
Then we have $\gcd(U'(2), 2^{2p}+1)=\gcd(U'(2), 2^{10}+1)=\gcd(U'(2), 25\cdot 41)=5.$
\end{proof}

\textbf{Theorem 3.5}
For $\mathbf{b}=(b(0), b(1), b(2), b(3))=(0, 1, 0, 1)$ or (1, 0, 1, 0), the 2-adic complexity of the sequence u defined in Lemma 2.1 is

$$\Phi_2(u)=\log_2\frac{2^{4p}-1}{5}.$$

\begin{proof}
We need to determine $\Phi_2(u')$ only.
From the definition of the 2-adic complexity, we have $\Phi_2(u')=\log_2\frac{2^{4p}-1}{\gcd(2^{4p}-1, U'(2))}.$ Since $\gcd(2^{2p}+1, 2^{2p}-1)=1,$ we know $\Phi_2(u')=\log_2\frac{2^{4p}-1}{\gcd(2^{2p}+1, U'(2)) \gcd(2^{2p}-1, U'(2))}.$
From Lemma 3.3 and Theorem 3.4, we get
$$\Phi_2(u')=\log_2\frac{2^{4p}-1}{\gcd(2^{2p}+1, U'(2))\gcd(2^{2p}-1, U'(2))}=\log_2\frac{2^{4p}-1}{5}.$$
\end{proof}

In the following, we will determine the 2-adic complexity of $u''$, the following two Lemmas are useful.

\textbf{Lemma 3.6} (\cite{H. G. Hu, Y. Sun3}) Let $U(x)$ and $T(x)$ be defined in Section 2. Then for a binary sequence $u$ with period $N$, we have

$$-2U(x)T(x^{-1})\equiv N+\sum_{\tau=1}^{N-1}C_u(\tau)x^\tau-T(x^{-1})(\sum_{i=0}^{N-1}x^i)\pmod{x^N-1}.$$

\textbf{Lemma 3.7} \cite{W. Su}
Let $\tau=\tau_1+4\tau_2$, where $\tau_1=0, 1\leq \tau_2\leq p-1$ or $1\leq \tau_1\leq 3, 0\leq \tau_2\leq p-1$. Then the autocorrelation function of $u''$ is
\begin{align*}
  C_{u''}(\tau)
 = \left\{ \begin{array}{ll}
  -4, &\textrm{$\tau_1=0, \tau_2\neq0,$}\\
  4,  &\textrm{$\tau_1=1, \tau_2+d\equiv 0\pmod p,$}\\
  4y, & \textrm{$\tau_1=1, \tau_2+d\pmod p\in D_0\cup D_2,$}\\
  -4y, & \textrm{$\tau_1=1, \tau_2+d\pmod p\in D_1\cup D_3,$}\\
  4, & \textrm{$\tau_1=2, \tau_2+2d\equiv 0\pmod p,$}\\
  0, & \textrm{$\tau_1=2, \tau_2+2d \not \equiv 0\pmod p,$}\\
 4, & \textrm{$\tau_1=3, \tau_2+3d \equiv 0\pmod p,$}\\
-4y, & \textrm{$\tau_1=3, \tau_2+3d\pmod p\in D_0\cup D_2,$}\\
 4y, &  \textrm{$\tau_1=3, \tau_2+3d\pmod p\in D_1\cup D_3.$}
\end{array} \right.
\end{align*}

\par\noindent\textbf{Lemma 3.8}
Let the symbols be the same as before. Then
 \begin{align*}
& U''(2)T''{(2^{-1})}\\
& \equiv2  \left[ \frac{2^{4p}-1}{2^{4}-1}-(2^{2p}+1)(2^p+1)+2^p(2^{2p}-1)y\sum_{i\in\mathbb{F}_p^{\ast}}(\frac{i}{p})2^{4i}-p \right]\pmod{2^{4p}-1}.
\end{align*}

\begin{proof}
From $-d\equiv\frac{p-1}{4}\pmod p$ and Lemma 3.7,
we have \begin{align*}
 &\sum_{\tau=1}^{4p-1}C_{u''}(\tau)2^{4\tau}\\
 & = \sum_{\tau_2=1}^{p-1}C_{u''}(4\tau_2)2^{4\tau_2}+\sum_{\tau_1=1}^3\sum_{\tau_2=0}^{p-1}C_{u''}(\tau_1+4\tau_2)2^{\tau_1+4\tau_2}\\
 & = -4\sum_{\tau_2=1}^{p-1}2^{4\tau_2}+4\cdot 2^{1+4\cdot\frac{p-1}{4}}+4y\sum_{\tau_2\in (D_0\cup D_2)+\frac{p-1}{4}}2^{1+4\tau_2}
-4y\sum_{\tau_2\in (D_1\cup D_3)+\frac{p-1}{4}}2^{1+4\tau_2}\\& \ \ \ +4\cdot2^{2+4\cdot\frac{p-1}{2}}+4\cdot2^{3+4\cdot\frac{3(p-1)}{4}}
                                  -4y\sum_{\tau_2\in (D_0\cup D_2)+\frac{3(p-1)}{4}}2^{3+4\tau_2}+4y\sum_{\tau_2\in (D_1\cup D_3)+\frac{3(p-1)}{4}}2^{3+4\tau_2}\\
                                  & \equiv -4\left[ \ \frac{2^{4p}-1}{2^4-1}-(1+2^{2p})(2^p+1)-2^py\sum_{i\in\mathbb{F}_p^{\ast}}(\frac{i}{p}) 2^{4i}+2^{3p}y\sum_{i\in \mathbb{F}_p^{\ast}}(\frac{i}{p})2^{4i}\right]\pmod{2^{4p}-1}.
\end{align*}
From Lemma 3.6 we get
\begin{align*}
& U''(2)T''{(2^{-1})}\\
& \equiv2  \left[ \frac{2^{4p}-1}{2^{4}-1}-(2^{2p}+1)(2^p+1)+2^p(2^{2p}-1)y\sum_{i\in\mathbb{F}_p^{\ast}}(\frac{i}{p})2^{4i}-p \right]\pmod{2^{4p}-1}.
\end{align*}
\end{proof}
\par\noindent\textbf{Lemma 3.9}
$\gcd(U''(2), 2^{2p}-1)=3.$
\begin{proof}
From Lemma 3.8 we know \begin{align*}
U''(2)T''({2^{-1}})\equiv &2[-(1+2^{2p})(2^p+1)-p]\pmod {\frac{2^{2p}-1}{3}}\\
                                   \equiv & 2[-2(2^p+1)-p]\pmod {\frac{2^{2p}-1}{3}}.
\end{align*}
Then $U''(2)T''(2^{-1})\equiv 2(-4-p)\pmod {2^p-1}$ and $U''(2)T''(2^{-1})\equiv -2p\pmod {\frac{2^{p}+1}{3}}.$

(1). We prove $\gcd(U''(2), 2^p-1)=1$ firstly. Let $l_1$ be a prime divisor of $\gcd(2^p-1, -4-p)$. Then $2^p\equiv 1\pmod {l_1}.$ From Fermat's
theorem, we know that $p|l_1-1$ which contradicts to $l_1|-p-4.$ Therefore
$\gcd(U''(2) T''(2^{-1}), 2^p-1)=\gcd(-4-p, 2^p-1)=1$ which implies that $\gcd(U''(2), 2^p-1)=1.$

(2). Next we prove that $\gcd(U''(2), \frac{2^p+1}{3})=1.$ Suppose that $l$ is a common prime divisor of $U''(2)$ and $\frac{2^p+1}{3}$. Then, by Lemma 3.8, $0\equiv U''(2)T''(2^{-1})\equiv -2p\pmod l$
so that $l=p$. From $-1\equiv 2^p\equiv 2\pmod p$ we get $p=3$ which contradicts to $p\equiv 1\pmod 4$. Therefore $\gcd(U''(2), \frac{2^p+1}{3})=1.$

(3). At last, we prove $3|U''(2)$.
By the definition of $U''(2)$, we get
\begin{align*}
U''(2)&=\sum_{i\in D_1\cup D_2}2^{4i}+\sum_{i\in D_0\cup D_3}2^{4(i+\frac{p-1}{4})+1}+\sum_{i\in D_0\cup D_1}2^{4(i+\frac{p-1}{2})+2}+\sum_{i\in D_0\cup D_1}2^{4(i+\frac{3(p-1)}{4})+3}\\
                                   & =\sum_{i\in D_1\cup D_2}2^{4i}+2^p\sum_{i\in D_0\cup D_3}2^{4i}+2^{2p}\sum_{i\in D_0\cup D_1}2^{4i}+2^{3p}\sum_{i\in D_0\cup D_1}2^{4i}\\
                                   & \equiv \frac{p-1}{2}- \frac{p-1}{2}+ \frac{p-1}{2}- \frac{p-1}{2} \pmod 3\\
                                   & \equiv 0 \pmod 3.
\end{align*}
From (1)-(3) we get $$\gcd(U''(2), 2^{2p}-1)=3\cdot\gcd \left(\frac{U''(2)}{3}, \frac{2^p+1}{3}\right)\cdot \gcd\left(U''(2), 2^p-1\right)=3.$$
\end{proof}
\par\noindent\textbf{Lemma 3.10}
$\gcd(U''(2), \frac{2^{2p}+1}{5})=1$ for $p\neq 5$.

The proof of this lemma is similar to Theorem 3.4, we omit it.
\par\noindent\textbf{Lemma 3.11}
$\gcd(U''(2), 2^{2p}+1)=25$ for $p=5.$

\begin{proof}
From $\mathbb{F}_5^{\ast}=\langle2\rangle$, we know the four cyclotomic classes of order four are $D_0=\{1\}$, $D_1=\{2\}$, $D_2=\{4\}$ and
$D_3=\{3\}$. For a binary periodic sequence $s$, we have $B_{L^d(s)}=B_s-d$. From the definition of $u''$ we have

\begin{align*}
 U''(2)& = \sum_{i\in D_1\cup D_2}2^{4i}+\sum_{i\in D_0\cup D_3}2^{4(i+\frac{p-1}{4})+1}+\sum_{i\in D_0\cup D_1}2^{4(i+\frac{p-1}{2})+2}+\sum_{i\in D_0\cup D_1}2^{4(i+\frac{3(p-1)}{4})+3}\\
   & = \sum_{i\in\{2, 4\}}2^{4i}+\sum_{i\in\{2, 4\}}2^{4i+1}+\sum_{i\in\{3, 4\}}2^{4i+2}+\sum_{i\in\{4, 5\}}2^{4i+3}\\
                                  & = 9388800\\
                                  & \equiv\left\{ \begin{array}{ll}
0, \pmod {25},\\
5, \pmod {41}.
\end{array} \right.
\end{align*}
Then we have $25|U''(2)$ and then from $2^{2p}+1=5^2\times 41$, we get $\gcd(U''(2), 2^{2p}+1)=25.$
\end{proof}
\textbf{Theorem 3.12} For $\mathbf{b}=(b(0), b(1), b(2), b(3))=(0, 0, 0, 0)$ or (1, 1, 1, 1), the 2-adic complexity of the sequence u defined in Lemma 2.1 is
\begin{align*}
\Phi_2(u)=\left\{ \begin{array}{ll}
\log_2\frac{2^{4p}-1}{75}, \textrm{$p=5$}\\
\log_2\frac{2^{4p}-1}{15}, \textrm{$p\neq5$ }.
\end{array} \right.
\end{align*}

\begin{proof}
From $p\equiv 1\pmod 4$ and $2^4\equiv 1\pmod 5$ we get $2^p\equiv 2\pmod 5.$ Then by the definition of $u''$,
\begin{align*}
U''(2)&=\sum_{i\in D_1\cup D_2}2^{4i}+\sum_{i\in D_0\cup D_3}2^{4(i+\frac{p-1}{4})+1}+\sum_{i\in D_0\cup D_1}2^{4(i+\frac{p-1}{2})+2}+\sum_{i\in D_0\cup D_1}2^{4(i+\frac{3(p-1)}{4})+3}\\
                                   & \equiv \sum_{i\in D_0\cup D_2}1+\sum_{i\in D_0\cup D_3}2+\sum_{i\in D_0\cup D_1}(4+8)\equiv 15\cdot \frac{p-1}{2}\equiv 0\pmod 5.
\end{align*}
If $p\neq5,$ by Lemma 3.9 and 3.10 we get $\Phi_2(u'')=\log_2(\frac{2^{4p}-1}{C})$ where
$$D=\gcd(U''(2), 2^{4p}-1)=5\cdot\gcd \left(\frac{U''(2)}{5}, \frac{2^{2p}+1}{5}\right)\cdot \gcd\left(U''(2), 2^p-1\right)=15.$$
For $p=5$, by Lemma 3.11 and 3.9 we get $C=\gcd(U''(2), 2^{2p}+1)\cdot\gcd(U''(2), 2^{2p}-1)=75.$
\end{proof}

At the end of this section we give an example to illustrate our main results.
\par\noindent\textbf{Example} Let $q=13=3^2+4\cdot 1^2$, $\mathbb{F}_{13}^{\ast}=\langle2\rangle$. The cyclotomic classes of order 4 in $\mathbb{F}_{13}$ are $D_0=\{1, 3, 9\}$, $D_1=\{2, 5, 6\}$,
$D_2=\{4, 10, 12\}$, $D_3=\{7, 8, 11\}$. Let $\mathbf{b}=(b(0), b(1), b(2), b(3))=(0, 1, 0, 1).$ From the definition of the sequence $u$ in Lemma 2.1, we have
\begin{align*}
 U'(2)& = \sum_{i\in D_1\cup D_2}2^{4i}+\sum_{i\in\{\frac{p-1}{4}\}\cup((D_1\cup D_2)+\frac{p-1}{4})}2^{4i+1}\\& \ \ \
   +\sum_{i\in (D_0\cup D_1+\frac{p-1}{2})}2^{4i+2}+\sum_{i\in\{\frac{3(p-1)}{4}\}\cup ((D_2\cup D_3)+\frac{3(p-1)}{4})}2^{4i+3}\\
                                  & = \sum_{i\in\{2, 5, 6, 4, 10, 12\}}2^{4i}+\sum_{i\in\{3, 5, 8, 9, 7, 0, 2\}}2^{4i+1}+\sum_{i\in\{7, 9, 2, 8, 11, 12\}}2^{4i+2}+\sum_{i\in\{9, 0, 6, 8, 3, 4, 7\}}2^{4i+3}.
\end{align*}
Let $\mathbf{b}=(b(0), b(1), b(2), b(3))=(0, 0, 0, 0).$ From the definition of the sequence $u$ in Lemma 2.1, we have
\begin{align*}
 U''(2)& = \sum_{i\in D_1\cup D_2}2^{4i}+\sum_{i\in D_0\cup D_3}2^{4(i+\frac{p-1}{4})+1}+\sum_{i\in D_0\cup D_1}2^{4(i+\frac{p-1}{2})+2}+\sum_{i\in D_0\cup D_1}2^{4(i+\frac{3(p-1)}{4})+3}\\
   & = \sum_{i\in D_1\cup D_2}2^{4i}+\sum_{i\in D_0\cup D_3}2^{4(i+3)+1}+\sum_{i\in D_0\cup D_1}2^{4(i+6)+2}+\sum_{i\in D_0\cup D_1}2^{4(i+9)+3}\\
   & = \sum_{i\in \{2, 4, 5, 6, 10, 12\}}2^{4i}+\sum_{i\in \{1, 3, 9, 7, 8, 11\}}2^{4(i+3)+1}+\sum_{i\in \{1, 3, 9, 2, 5, 6\}}2^{4(i+6)+2}\\& \ \ \
   +\sum_{i\in\{1, 3, 9, 2, 5, 6\}}2^{4(i+9)+3}.
\end{align*}
Computing with magma, we have $\gcd(U'(2), 2^{52}-1)=5$ and $\gcd(U''(2), 2^{52}-1)=15$. Then we get $\Phi_2(u')=\log_2\frac{2^{4p-1}}{5}$ and $\Phi_2(u'')=\log_2\frac{2^{4p-1}}{15}$ which coincides with Theorem 3.5 and 3.12, respectively.

\end{document}